\documentclass{elsart}         
\usepackage{natbib} \usepackage{epsfig}
\begin{document}
\begin{frontmatter}
\title{$e^+e^-$ annihilation at low energies in analytic approach to QCD}
\author{D.V. Shirkov\thanksref{DVS}}
\author{I.L. Solovtsov\thanksref{ILS}}
\address{Bogoliubov Laboratory of Theoretical Physics,
Joint Institute for Nuclear Research,
Dubna, Moscow Region, 141980, Russia}
\thanks[DVS]{e-mail: shirkovd@thsun1.jinr.ru}
\thanks[ILS]{e-mail: solovtso@thsun1.jinr.ru}

\begin{abstract}
We begin with short review of the analytic approach (AA) to QCD recently
developed and applied to the process of $e^+e^-$ annihilation into hadrons
at low energies.
Besides summary of the theoretical description of smeared experimental data
for the $R$ cross-section ratio we give fresh analogous result for the
corresponding Adler function, $D(Q^2)$, and demonstrate excellent agreement
between the AA theoretical results and data in the low $Q^2$ region.
\end{abstract}

\end{frontmatter}

{\bf 1.} The {\sf Invariant Analytization} (=analytic approach) in the
perturbative quantum chromodynamics (pQCD) based\footnote{This central idea
of Invariant Analytization -- combining RG invariance with the
K\"all$\acute{{\rm e}}$n-Lehmann analyticity in  $Q^2$ has been borrowed from
 QED of Ref.~\cite{blsh59}.} on the $Q^2$-analyticity has been proposed in
Ref.\cite{jinr96}. For the invariant (running) coupling it yields an analytic
expression $\bar{\alpha}_{\rm an}(Q^2)$ with unphysical singularities
subtracted by nonperturbative contribution so that its perturbative expansion
precisely coincides with the usual perturbation one.  An infrared (IR)
limiting value of this analytic coupling $\bar{\alpha}_{\rm
an}(0)=4\pi/\beta_0$ is finite and independent of the scale parameter
$\Lambda$. Quite remarkably, this value turns out to be insensitive, that is
stable, with respect to higher loop corrections. The Invariant Analytization
(IA) introduces no extra parameters, however, its results for running
coupling and observables seem to correlate different experimental data ---
see Ref.~\cite{prl97}.

According to Refs.~\cite{jinr96,prl97} the analytic coupling is defined by a
special ansatz~-- via the spectral integral with a spectral function
$\rho(\sigma)$ defined ``perturbatively" as a discontinuity of the usual,
renormalization group (RG) summed, invariant coupling continued analytically
on the physical cut. The one-loop expression for the analytic coupling thus
defined can be presented explicitly
\begin{equation}  \label{an1}
{a}_{\rm an}^{(1)}(Q^2)=\frac{\bar{\alpha}_{\rm an}(Q^2)}{\pi}
=\frac{1}{\beta_0} \left[\frac{1}
{\ln Q^2/\Lambda^2}\,+ \,\frac{\Lambda^2} {\Lambda^2-Q^2}\right]\, .
\end{equation}
Here, the ghost pole at $Q^2=\Lambda^2$ is removed by the non-perturbative
term which appears due to the spectral condition and is invisible in the
perturbative expansion. The same IA procedure, being applied to invariant
QCD coupling in two- and three-loop cases, produces more complicates
expressions with the same basic properties -- see, e.g.,
Refs.~\cite{jinr96,prl97,qcd97}. Their analysis has revealed an important
feature -- reasonable stability of the analytic coupling behavior in the
whole IR region with respect to higher loop corrections. The maximal
deviation of the two-loop curve of the one-loop expression (\ref{an1}) is
about 10 per cent at $\sqrt{Q^2} \simeq \Lambda/4$. A much smaller, within
one per cent, difference between two- and three-loop
${\overline{\rm MS}}$-scheme curves provided, in turn, a basis for
stability with respect to the renormalization scheme (RS) dependence.

{\bf 2.} Further on, the analytization idea has been applied to an analysis
of observables. The most simple possibility is to use the analytic
$\bar{\alpha}_{\rm an}(Q^2)$ expression instead of the usual
$\bar{\alpha}_s(Q^2)$ one with the ghost singularities. \par Meanwhile, the
Invariant Analytization of a physical amplitude $F(Q,\alpha)$ is not a
straightforward procedure. A few different scenarios are possible. In the
paper \cite{MSS97}, a particular version, the {\sf Analytic Perturbation
Theory} (APT), has been proposed and elaborated. Here, due to the specific
analytization ansatz, instead of the power perturbation series an analytic
amplitude ${\mathcal F}(x)$ is presented in a form of a more general
expansion over an asymptotic set of functions
${\mathcal A}_n(x)=\left[a^n(x) \right]_{\rm an}\,$,
the ``n-th power of $a(x)$
analytized as a whole". Analytization changes the nature of perturbation
expansion transforming it (see Ref.~\cite{tmp99}) into a non-power
asymptotic expansion a la Erdelyi.

 In the APT approach, the drastic reduction of loop and RS sensitivity for
several observables has been found -- see Refs.~\cite{MSS97,MSS97b,pl97}.
Now, we are going to concentrate on the item of the renormalization scheme
dependence which is common in the pQCD description of physical quantities due
to a truncation of a perturbation series.

{\bf 3.} Consider $R(s)$, the well known cross-section ratio for {\sf the
process of $e^+e^-$ annihilation into hadrons}. This example is physically
interesting for the RS stability issue (see, e.g., discussion in
Ref.~\cite{racz} and references therein). We first review shortly our
results of Ref.~\cite{pl97}.

The RS-invariant cross-section ratio, $R(s)$, has the form
$$R(s)=3\sum_{f}^{}\,Q_{f}^2\,[1+r_{f}(s)]\; , \ \
r_{f}(s)=\tilde{a}(s)\left[1+r_1\tilde{a}(s)+r_2\tilde{a}^2(s)\right], $$
where we wrote the QCD contribution, $r_{f}(s)$, in the third order for the
massless case. The function $\tilde{a}(s)$ and coefficients $r_k$ depend of
the flavor number $f$.

The function $\tilde{a}(s)$, ``the running coupling in the time-like
region", usually, is defined naively as a mirror image $\tilde{a}(-Q^2)
=a(Q^2)$. Instead, we use the following self-consistent with analyticity
relations
\begin{equation} \label{dipole}
d_{f}(Q^2)=Q^2\int_0^\infty\frac{ds}{(s+Q^2)^2}\,r_{f}(s)\, , \quad
r_{f}(s)= - \frac{1}{2\pi{\rm i}} \int^{s+{\rm i}\varepsilon}_{s-{\rm i}
\varepsilon} \frac{d\sigma}{\sigma}\, d_{f}(-\sigma)\, ,
\end{equation}
where the integration contour in the last expression lies in the region of
analyticity of $d_{f}(Q^2)$.

Expansion for the QCD correction $r(s)$ is similar to the one for the
RS-invariant Adler function $$ D(Q^2)=3\sum_{f}Q_{f}^2[1+d_{f}(Q^2)]\;, \ \
d_{f}(Q^2)=a(Q^2)\left[1+ d_1a(Q^2)+d_2a^2(Q^2)\right]\, .$$
For $d_{f}(Q^2)$ the K\"all$\acute{{\rm e}}$n--Lehmann representation with
some effective spectral function is valid. On the other hand, by using the
reverse relation for $r_{f}(s)$ in Eq.~(\ref{dipole}), one can express
$r_{f}(s)$ via the same spectral function (see Ref.~\cite{ms97}):
\begin{equation}  \label{d-spectral}
d_{f}(Q^2)=\frac{1}{\pi}\int_0^\infty\frac{d \sigma}{\sigma+Q^2}
\varrho^{\rm eff}_{f}(\sigma)\, ,\quad
r_{f}(s)=\frac{1}{\pi}\,\int_s^\infty\, \frac{d\,\sigma}{\sigma}
\,\varrho^{\rm eff}_{f}(\sigma)\, .
\end{equation}
The Adler spectral density $\varrho^{\rm eff}_{f}(\sigma)$ is represented
by an expansion
$\varrho^{\rm eff}_{f}(\sigma)
=\rho_0(\sigma)+d_1 \varrho_1(\sigma)+d_2\varrho_2(\sigma)$,
where the quantities in the r. h. s. depend on $f$. The first term,
$\rho_0(\sigma)$, in the last expression is just the spectral function for
$a(Q^2)$ and $\varrho_k(\sigma)={\rm Im}[a(-\sigma-{\rm i}\epsilon)]^{k+1}$
are related to its higher powers.
It is essential that the Adler function is defined in the Euclidean region
where the renormalization group method can be applied directly.

We will use the `cancellation index criterion' proposed in Ref.~\cite{racz95}
to qualify various RS's. According to this criterion, on the three-loop
level, a set of ``natural well-behaved schemes" can be defined.  The degree
of cancellation can be measured by the `cancellation index', $C_R$. To
discuss the RS-dependence, we consider two examples of ``well-behaved" RS's:
the scheme $H$ (the so-called 't Hooft scheme) with the parameters (for $f=3)
$ $r_1^{(H)}=-3.2$, $c_2^{(H)}=0$ and $\overline{\rm MS}$ scheme with
$r_1^{(\overline{\rm MS})}=1.64$, $c_2^{(\overline{\rm MS})}=4.47$.  Both the
schemes have a sufficiently small value $C_R\simeq2$, which is close to that
for the so-called optimal RS based on the principle of minimal sensitivity
(PMS), Ref.~\cite{stev}.

            \begin{figure}[hpt]
\centerline{\epsfig{file=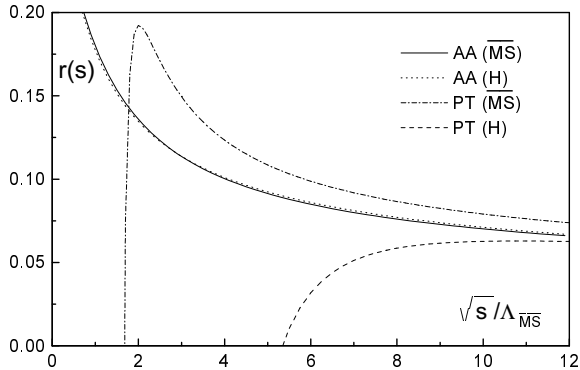,width=9.0cm} }
\caption{{\sl Plot of the QCD correction $r_{f=3}(s)$ calculated in the
cases of perturbation theory ({\rm PT}) and the analytic approach ({\rm AA})
in two different RS's with approximately the same cancellation index
$C_R\simeq 2$: H and $\overline{\rm MS}$. } } \label{RS-dep}
          \end{figure}

In Fig.~\ref{RS-dep}, we plot the QCD correction $r_{f=3}(s)$ as a function
of $\sqrt{s}/\Lambda_{\overline{\rm MS}}~$ for these two schemes in the
usual treatment, as it was considered, e.g., in Refs.~\cite{racz,ckl,ms} and
within the AA. One can see that the analytically improved result for $R(s)$
obeys a stable behavior for the whole interval of energies being practically
scheme-independent.

{\bf 4.} To incorporate threshold effects and compare our results with
experiment, we use  for the cross-section ratio  an approximate expression --
 see Ref.~\cite{PQW}
\begin{equation}  \label{R_approx}
\tilde{R}(s)=3\sum_f\, Q_f^2\, \Theta (s-4m_f^2)\, T(v_f)\,
\left[ 1\,+\,g(v_f) r_f(s) \right]\, , \end{equation}
$$v_f=\sqrt{1-4m_f^2/s}, \;  T(v)=v(3-v^2)/2, \;
g(v)=\frac{4\pi}{3}\left[\frac{\pi}{2v}-\frac{3+v}{4}
\left(\frac{\pi}{2}-\frac{3}{4\pi} \right) \right],$$
and consider the ``smeared" quantity
\begin{equation}  \label{R_smear}
R_{\Delta}(q^2)=\frac{\Delta}{\pi}\int_0^\infty\, ds
\frac{\tilde{R}(s)}{(s-q^2)^2+\Delta^2}\, .
\end{equation}

            \begin{figure}[htp]
\centerline{\epsfig{file=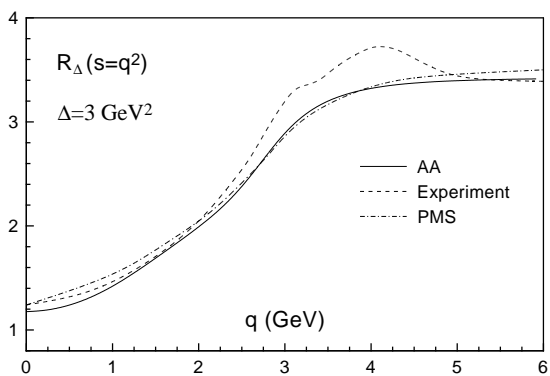,width=9.0cm}}
\caption{{\sl
The smeared $e^+e^-$ annihilation ratio $R_{\Delta}(q^2)$ vs. $q$.}}
\label{r_smear}
          \end{figure}

In Fig.~\ref{r_smear} we show smeared experimental data for
$R_{\Delta}(q^2)$ at $\Delta=3\,{\rm GeV}^2$ and the third-order PMS curve
taken from Ref.~\cite{ms}. In the same figure we plot our three-loop result
obtained with the value of the scale parameter as in Ref.~\cite{pl97}.

{\bf 5.}
Now we consider {\sf the Adler $D$-function}, for which new ``experimental"
data have been recently obtained in Ref.~\cite{EJKV98} and we will
compare our results with them.\footnote{Relevant topics have been discussed
on this Workshop in the talks given by F.~Jegerlehner and A.L.~Kataev.} It
should be noted that a comparison with the $D$-function data determined in
the Euclidean region does not require any ``smearing procedure" and can be
done directly.

\begin{figure}[htp]
\centerline{\epsfig{file=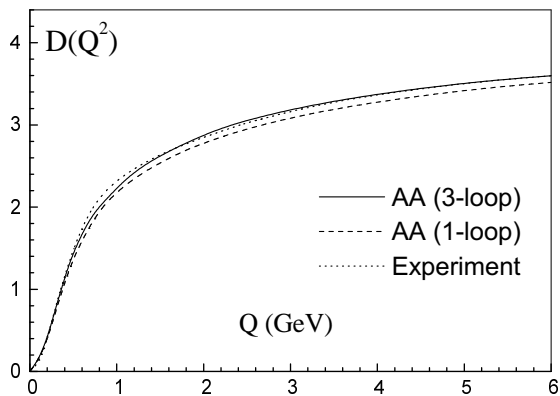,width=9.9cm}}
\caption{ {\sl The $D$-function vs. $Q$.  } }
\label{d-apt1}
\end{figure}

By using Eq.~(\ref{R_approx}), we can derive the Adler function, $D(Q^2)$.
Its low $Q$-shape is sensitive to the light quark mass values. To obtain the
best fit, these values have to be chosen near the constituent quark mass
values (compare with Ref.~\cite{jeg96}). Here, to derive $R_\Delta(s)$ and
$D(Q^2)$, we used $m_u=m_d=250$~MeV, $m_s=400$~MeV, $m_c=1.35$~GeV,
$m_b=4.75$~GeV, and $m_t=174$~GeV. In Fig.~\ref{d-apt1} we plot our fresh
results for $D$-function calculated via $r_{f}(s)$ in Eq.~(\ref{R_approx})
in the ${\overline{{\rm MS}}}$ scheme.  The three-loop curve agrees quite
well with the one of Ref.~\cite{EJKV98}. Note that our description of the
$D$-function is RS stable in accordance with general features of the AA.

{\bf 6.} The analytically improved running coupling being a smooth function
in the IR region turns out to be remarkably stable with respect to higher
loop corrections. Its shape agrees well with the IR integral characteristic
extracted from jet physics data. We have found further evidence that our AA
reduces the RS dependence drastically: $r_{\rm an}(s)$ thus obtained turns
out to be practically scheme-independent in a wide class of RS for the whole
energy interval. We have also demonstrated that the AA description agrees
with experimental data for the smeared $e^+e^-$ annihilation ratio and the
vector current $D$-function quite well.

The authors would like to thank A.L.~Kataev who brought the paper
\cite{EJKV98} to our attention. We also grateful to him and F.~Jegerlehner
for useful discussions of the results obtained.  The partial support of the
RFFI grants Nos. 96-15-96030, 99-01-00091, and INTAS 96-0842 is appreciated.

\end{document}